\documentclass[a4paper]{jpconf}
\usepackage{graphicx}
\def\ncphoton{$\gamma$--NCP}

\begin{document}
\title{Photon neutrino-production in a chiral EFT for nuclei and extrapolation to $E_{\nu}\sim$ GeV region}

\author{X Zhang}

\address{Institute of Nuclear and Particle Physics and Department of
Physics and Astronomy, Ohio University, Athens, OH 45701, USA }

\ead{zhangx4@ohio.edu}

\begin{abstract}
We carry out a series of studies on pion and photon productions in neutrino/electron/photon--nucleus scatterings. The low energy region is investigated by using a chiral effective field theory for nuclei. The results for the neutral current induced photon production (\ncphoton) are then extrapolated to neutrino energy $E_{\nu}\sim$ GeV region. By convoluting the cross sections with MiniBooNE's beam spectrum and detection efficiency, we estimate its \ncphoton\, event number, and conclude that such photon production can not fully explain its low energy event excess in both neutrino and antineutrino runs. 

\end{abstract}

\section{Introduction}
I briefly summarize our study of photon and pion productions in neutrino-nucleus scattering. The pion productions are used to benchmark the framework applied here: the charged current induced pion production from nucleons, for which there exists experimental data, is used to calibrate the interaction kernel composed of both $\Delta$ and non-resonant  contributions \cite{Serot:2010xn,Serot:2011yx,Serot:2012rd}; the incoherent electro-production \cite{Zhang:2012aka} and coherent photo-production \cite{Zhang:2012xi} are used to benchmark the kernel medium-modifications and the approximation schemes. 
We then apply the same ingredients to study \ncphoton. The interaction kernel is derived by using the same Lagrangian as used in pion production. At the low energy region, i.e., neutrino energy $E_{\nu} \leq 0.5$ GeV, we work in the so-called quantum hadrondynamics effective field theory (QHD-EFT) \cite{Serot:1997xg}, which is Lorentz covariant and includes nonlinear realization of chiral symmetry and (vector) meson dominance. The EFT has been used extensively to study medium-heavy nucleus structure and electromagnetic nuclear response \cite{Serot:1997xg}. We then extrapolate the results to $E_{\nu} \sim 1 $ GeV region by using phenomenological form factors in the interaction vertices and compute \ncphoton\, events in the MiniBooNE experiment \cite{Zhang:2012xn}, in order to address \ncphoton's role in its  low energy excess \cite{MiniBNoscsum}.

\section{Theory benchmarks}
\begin{figure}
\centering
\includegraphics[width=5cm,angle=-90]{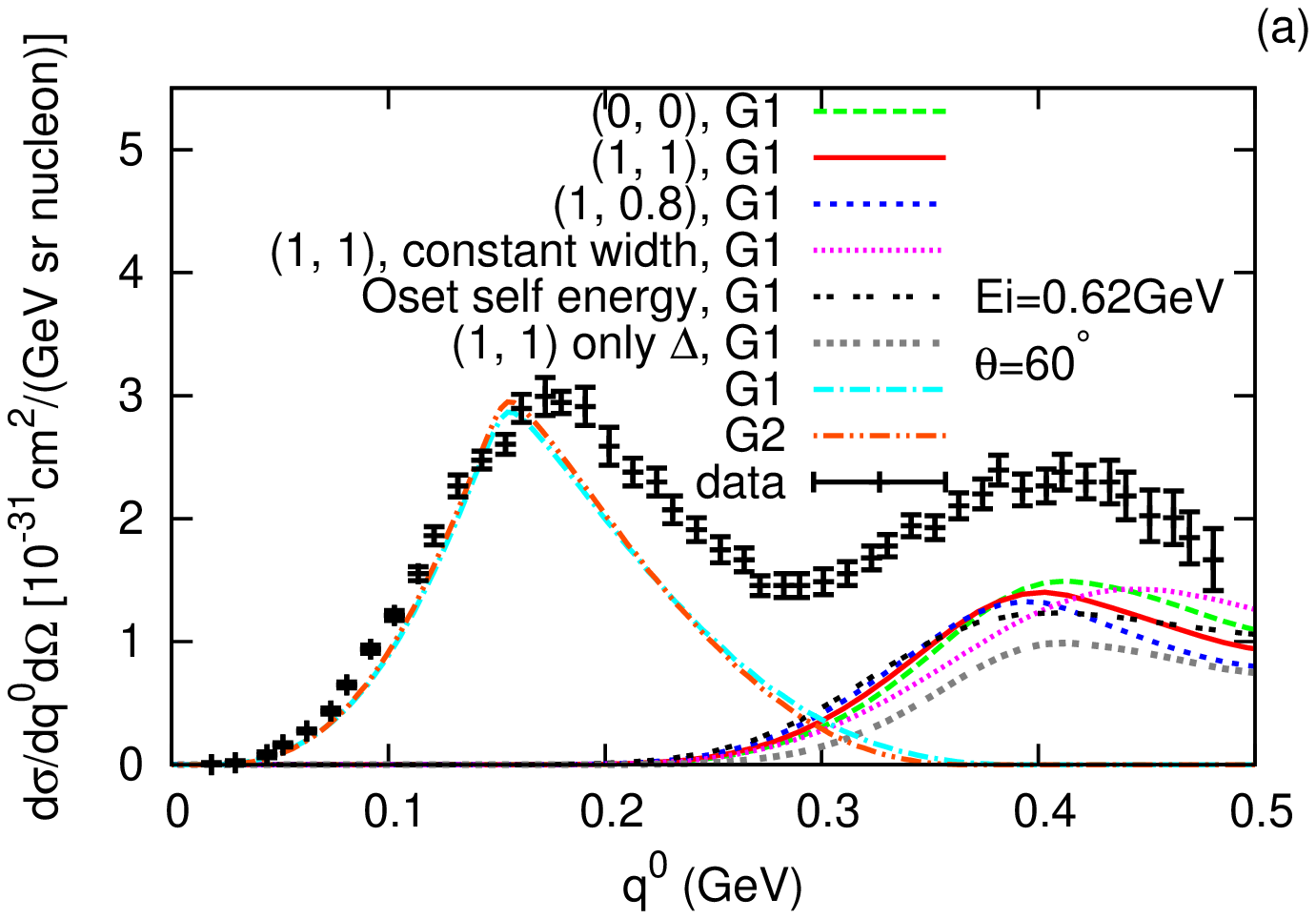}
\includegraphics[width=5cm,angle=-90]{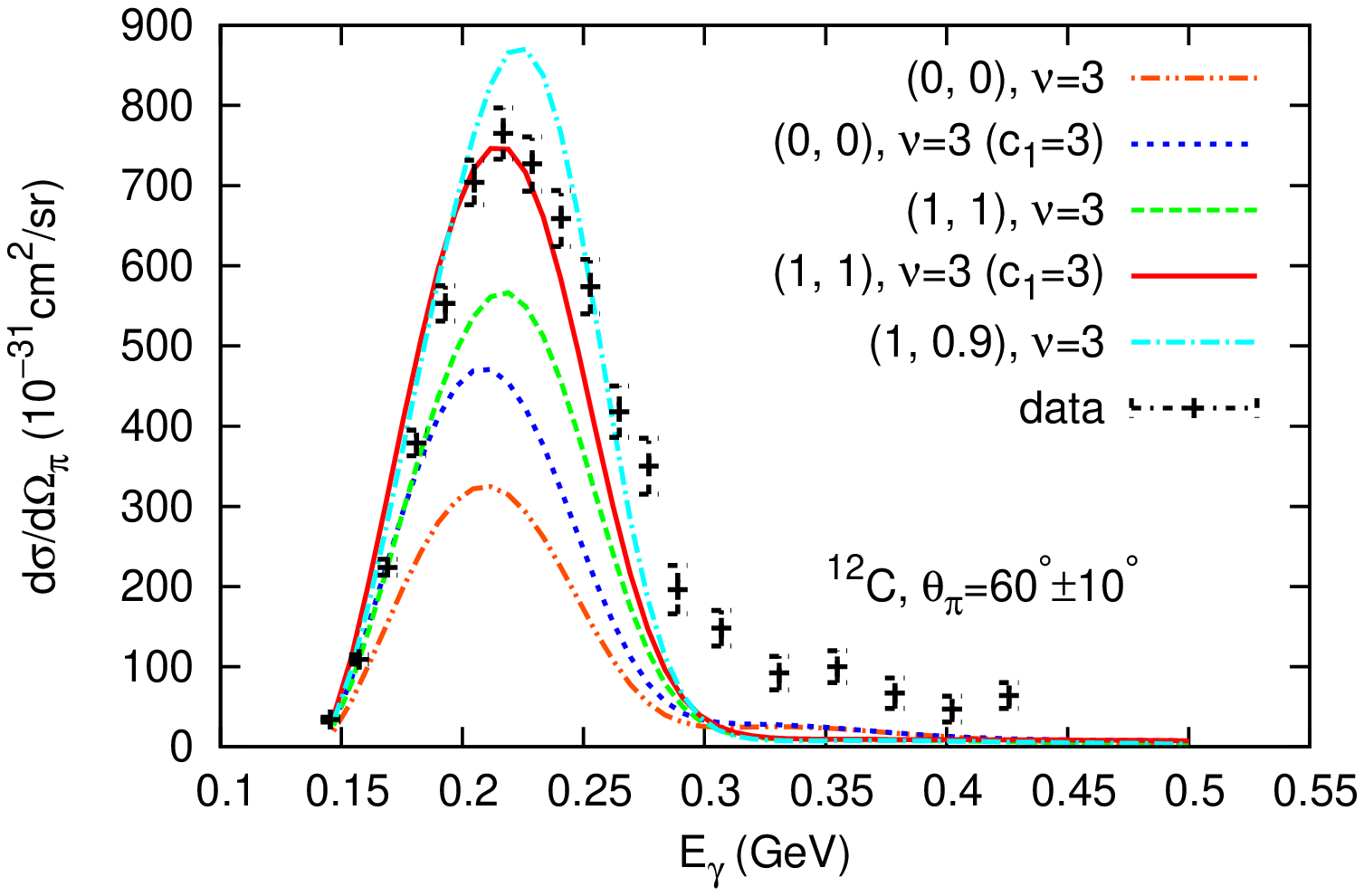}
\caption{The left panel shows the data \cite{Barreau83} for inclusive electron-${}^{12}C$ scattering differential cross section at different energy deposit with incoming electron  energy $E_{i}=0.62 \ \mathrm{GeV}$ and its scattering angle $\theta_{lf}=60^{\circ}$, and our quasi elastic scattering and incoherent pion production results. The right panel shows $d\sigma/d\Omega_{\pi}$ of the coherent $\pi^{0}$ photo-production from ${}^{12}C$ vs. the photon energy $E_{\gamma}$. The final pion angle is fixed at $\theta_{\pi}=60^{\circ}\pm10^{\circ}$. 
The data are from Ref.~\cite{Schmitz96}. The detailed discussions can be found in text. }
\label{fig:electroscattering}
\end{figure}
For a detailed account of the interaction kernel and QHD-EFT, please see Refs.~\cite{Serot:2010xn,Serot:2011yx,Serot:2012rd}. Here I focus on the kernel medium modifications. In the QHD-EFT, the strong interaction between nucleons ($N$) is due to  meson exchanges including isoscalar scalar $\phi$, isoscalar vector $V^{\mu}$, isovector vector $\rho^{i}_{\mu}$, and pion $\pi^{i}$ mesons. In the mean field approximation, $\phi$ and $V^{\mu}$ develop nonzero expectation values in isoscalar nucleus; the nucleon single particle spectrum is modified accordingly. This simple picture explains nucleon's large spin-orbital ($L$-$S$) coupling \cite{Serot:1997xg}. We introduce the same  couplings between $\Delta$ resonances and the two mesons \cite{Zhang:2012aka}, and get a $L$-$S$ coupling consistent with the phenomenological fit based on $\pi$-nucleus scattering when the ratios between ($\Delta$-$\phi$, $\Delta$-$V^{\mu}$)  and ($N$-$\phi$, $N$-$V^{\mu}$) couplings, denoted as ($r_{s}$, $r_{v}$) in the following, are around $(1,\, 1)$. On the other hand, the imaginary part of $\Delta$'s spectrum is modeled by adding a spreading potential to its pion decay width (in medium) \cite{Zhang:2012aka}, in contrast to a sophisticated nonrelativistic framework study \cite{oset87}. (It is certainly interesting to carry out similar study in the QHD-EFT.) As the result both nucleon and $\Delta$ spectrum are modified in the same framework.

 We then study the incoherent electron-nucleus scatterings, including the quasi-elastic and pion production (see the left panel of Fig.~\ref{fig:electroscattering}), and the coherent pion photo-production (see the right panel of Fig.~\ref{fig:electroscattering}). In both plots,  $(1,1)$ denotes $(r_{s}=1,r_{v}=1)$ and so do other $(\cdots,\cdots)$. In the left panel, the first three curves (from top to bottom in the legend) use our medium modifications with different $r_s$ and $r_v$; the fourth and fifth curves, labeled as ``constant width'' and ``Oset self energy'', are results by using a constant increase of $\Delta$ width and by the self-energy insertion given in Ref.~\cite{oset87}. The kernels of these five calculations include  both $\Delta$ and non-resonant contributions. ``(1,1) only $\Delta$'' includes only $\Delta$'s contribution in the kernel  with $(r_{s}=1,r_{v}=1)$. The last two curves correspond to the quasi-elastic channel with the ``G1'' and ``G2'' parameter sets (for describing the nuclear structure)\cite{Zhang:2012aka}.  Because they are almost the same, in the  following we only use the ``G1'' set. We clearly see that $r_s$ and $r_v$ dial the position of the $\Delta$ peak. Because only the quasi elastic and  pion production channels are included, it is not a surprise that the inclusive data are significantly above our results. We can add the two-body current contribution and turn off $\Delta$ width increase, and get a total cross section consistent with the inclusive data (see Ref.~\cite{Zhang:2012aka} for details). By using the same kernel and medium modifications, we study the coherent photo-production of $\pi^{0}$. The right panel of Fig.~\ref{fig:electroscattering} shows $d\sigma/d\Omega_{\pi}$ vs. photon energy at fixed pion angle $\theta_{\pi}=60^{\circ}$. Here $c_{1}$ is a coupling involving $Z$ boson (or $\pi$), photon, and nucleon. The plot clearly shows that the $\pi^{0}$ production is sensitive to this coupling and also to $(r_s,\, r_v)$. By comparing curves with different combinations of the three parameters, we see $(r_{s}=1,\, r_{v}=1)\, c_{1}=3$ and $(r_{s}=1,\, r_{v}=0.9)$
($c_{1}=1.5$ without being mentioned explicitly) agree with data reasonably well for small photon energy $E_{\gamma} \leq 0.3$ MeV, but all the curves systemically underestimate the differential cross section above that. (``$\nu=3$'' means full kernel.)
It should be pointed that the $c_{1}$ coupling was first proposed to enhance \ncphoton\, in order to explain MiniBooNE low energy excess (see Ref.~\cite{Hill10} and references therein). As will be demonstrated in the lower panel of Fig.~\ref{fig:nuxsection}, the coherent \ncphoton\, is indeed  sensitive to these coupling.  So we  use these two parameter sets, $(r_{s}=1,\, r_{v}=1),\, c_{1}=3$ and $(r_{s}=1,\, r_{v}=0.9),\, c_{1}=1.5$, to estimate the uncertainty of \ncphoton\, cross section. It should be emphasized that in the low energy incoherent productions (e.g. electro-production shown previously), $c_{1}$ contribution can be ignored \cite{Serot:2012rd}. 

\section{\ncphoton\, at GeV region and at MiniBooNE}
\begin{figure}
\centering
\includegraphics[width=4.1cm, height=6.3cm, angle=-90] {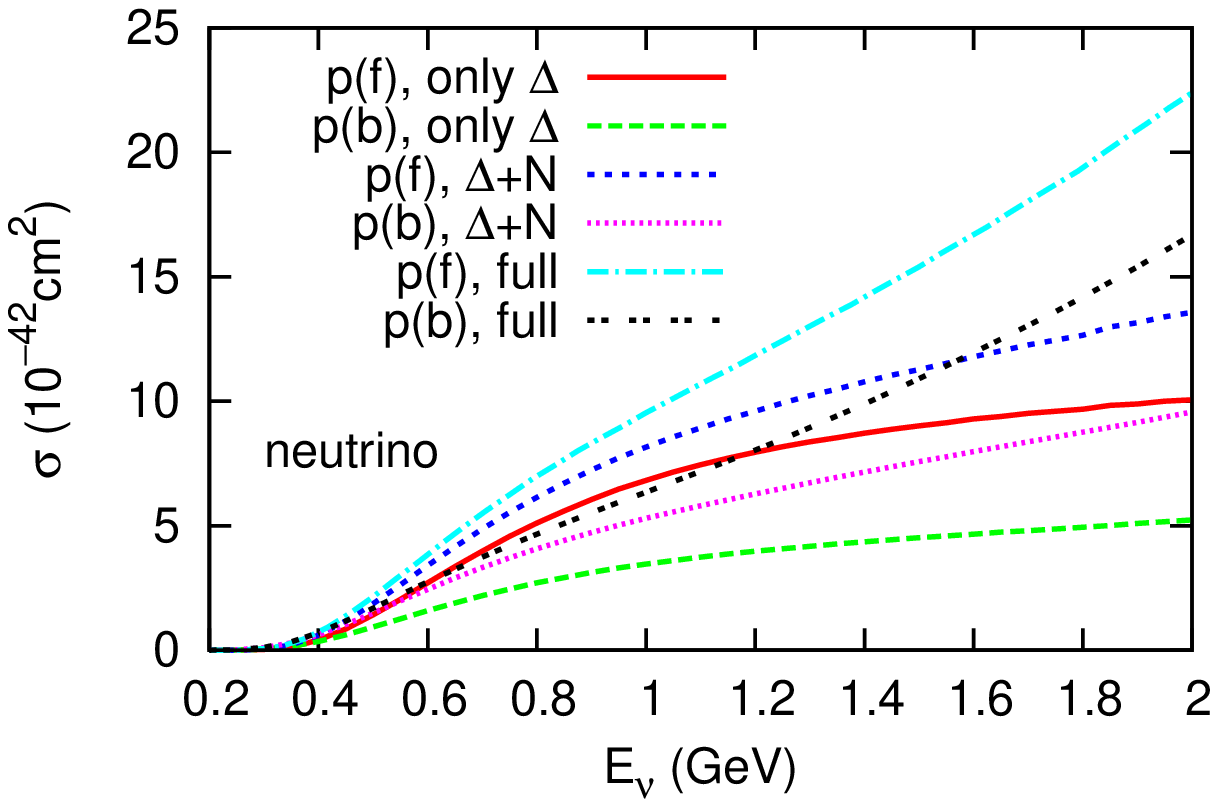}
\includegraphics[width=4.1cm, height=6.3cm,angle=-90] {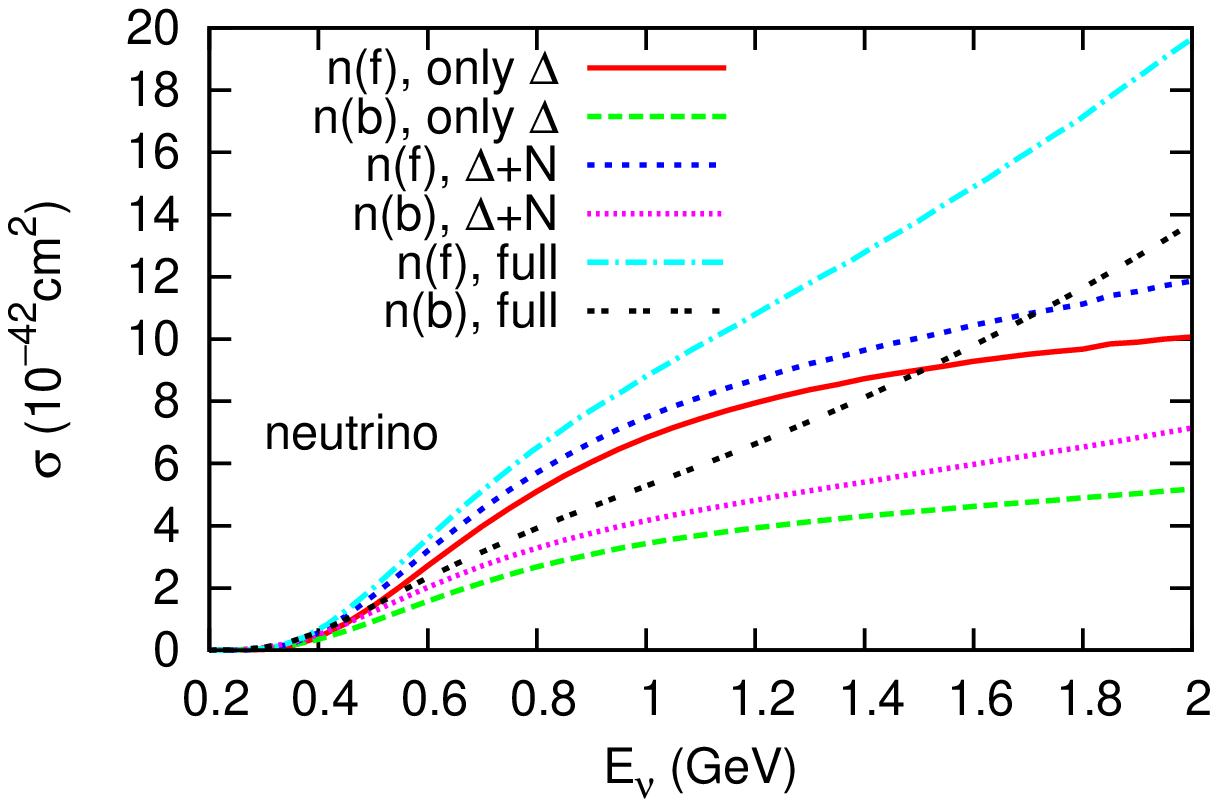}
\includegraphics[width=4.1cm, height=6.3cm, angle=-90] {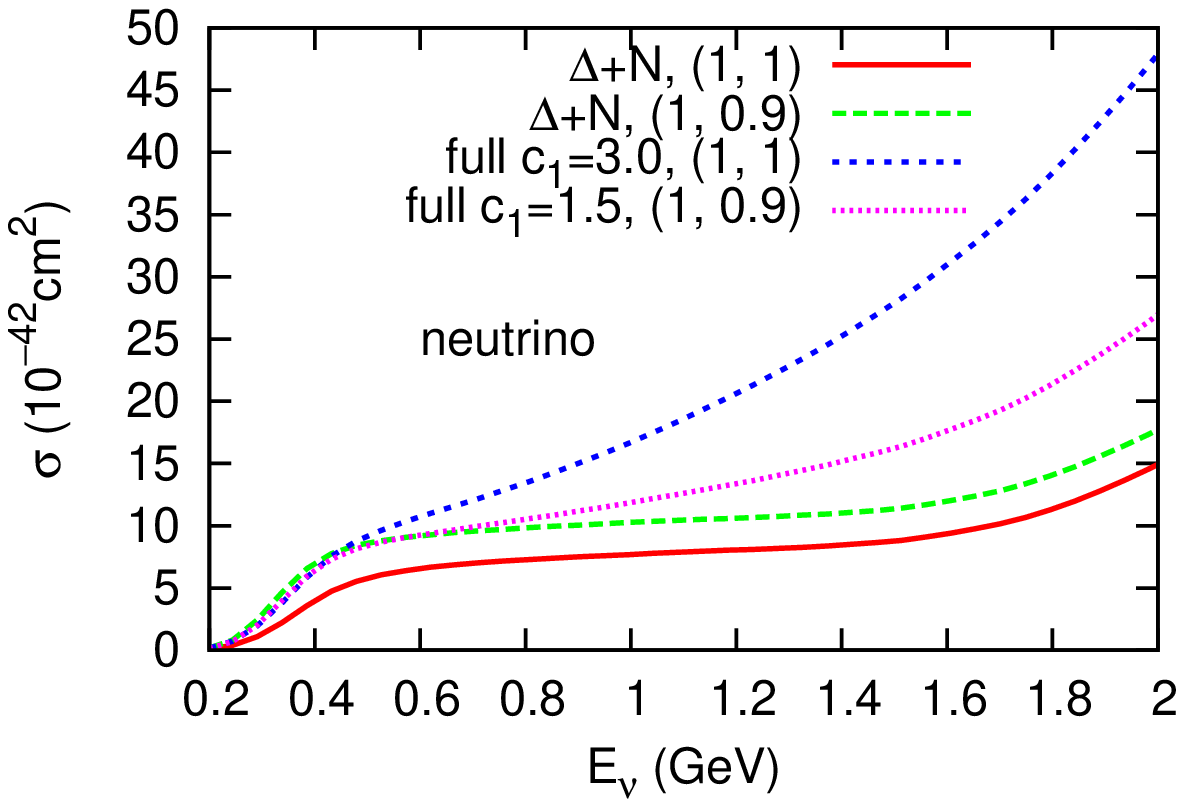}
\caption{The upper left panel shows the total cross section of \ncphoton\, in neutrino-free-proton scattering (i.e.,``p(f)'') and the incoherent production in neutrino-${}^{12}C$ scattering (on bounded proton, i.e., ``p(b)''). In parallel, the right is for the productions from neutrons. The lower panel shows the coherent production cross section from ${}^{12}C$. See the text for the detailed discussion.}
\label{fig:nuxsection}
\end{figure}
Fig.~\ref{fig:nuxsection} upper left panel shows the \ncphoton\, cross sections in neutrino-free-proton [denoted as ``p(f)''] and neutrino-bounded-proton (in ${}^{12}C$) scatterings [denoted as ``p(b)''], while those for neutron scatterings are shown in the right panel. Results for antineutrino scatterings can be found in Ref.~\cite{Zhang:2012xn}. In each plot, three different kernels are used: ``only $\Delta$'' has $\Delta$ contribution; ``$\Delta+N$'' includes both $\Delta$ and nucleon intermediate state contributions; and ``full'' includes all the diagrams including the mentioned $c_1$ coupling \cite{Zhang:2012xn}. Here we use $(r_s=1,\,r_v=1),\, c_{1}=3$; its difference from using $(r_{s}=1,\, r_{v}=0.9),\, c_{1}=1.5$ is small when $E_{\nu}\sim 1$ GeV. We can see $\Delta$ dominates in the $1$ GeV region. The large non-resonant contribution above $1$ GeV should be taken cautiously, because its regularization is model dependent. In the lower panel of Fig.~\ref{fig:nuxsection}, we show the total cross section for coherent \ncphoton\, in neutrino-${}^{12}C$ scattering. The first two curves include $\Delta$ and nucleon intermediate state contributions, and the other two implement the full interaction kernel with $(r_s,\,r_v),\, c_1$ chosen to reproduce the coherent pion photo-production data (see Fig.~\ref{fig:electroscattering}). 
It is clearly demonstrated here that the coherent production is sensitive to the three parameters. Again the non-resonant contribution above $1$ GeV depends on the regularization. 

However, in MiniBooNE the median beam energy is around $0.5-1$ GeV, so the large cross section uncertainty above $1$ GeV does not necessarily lead to big uncertainty in \ncphoton\, event estimate. The Table~\ref{tab:summary_EQE} shows our results in different $E_{QE}$ bins for coherent and incoherent productions, and production from free protons, based on MiniBooNE beam spectrum and its detector ($\mathrm{CH}_{2}$) efficiency \cite{Zhang:2012xn}.  $E_{QE}$ is the reconstructed neutrino energy \cite{MiniBNoscsum} which can be different from the true neutrino energy because the final photon can take away significant energy.  In each entry, we show the lower bound and upper bound based on the biggest and smallest ``full'' cross section results shown in Fig.\ref{fig:nuxsection}. 
The same table for MiniBoNE's antineutrino run can be found in Ref.~\cite{Zhang:2012xn}. {\it Based on this, we conclude that the difference between our result and MiniBooNE's is not large enough to explain all the low energy excess!} The same conclusion has been claimed by the other study \cite{Alvarez-Ruso:2013ica} (see WG2 talk by En Wang in this workshop ), but the difference between the two calculations' incoherent productions needs to be resolved in the future.

It should be noted that a proposal \cite{Dharmapalan:2013zcy} has been made to add scintillator to the MiniBooNE detector to test CC and/or NC origin of its low energy excess. In addition, ArgoNeuT and future MicroBooNE experiments can also identify \ncphoton\, event (see WG2 talk by Tingjun Yang in this workshop).
On the theoretical side, a better understanding of the kernel in the resonance region can be achieved by using dispersion relation formalism, which is currently being studied. Moreover, combining photon and pion productions can help disentangle the final state interaction from the kernel (and formation time/zone effect \cite{Golan:2012wx}) in various neutrino-nucleus scattering modelings \cite{Golan:2012wx}, if the experimental measurements can be precise enough.

\begin{table} 
  \begin{center}
    \begin{tabular}{|c|c|c|c|} \hline
$E_{QE}(\mathrm{GeV})$   & $[0.2 \, , \, 0.3]$ 
                  & $[0.3 \, , \, 0.475]$
                  & $[0.475 \, , \, 1.25]$ \\[2pt] \hline
$\mathrm{coh}$    & $1.5\ (2.9)$
                  & $6.0\ (9.2)$
                  & $2.1\ (8.0)$  \\[2pt] \hline
$\mathrm{inc}$  & $12.0\ (14.1)$
                  & $25.5\ (31.1)$
                  & $12.6 \ (23.2)$  \\[2pt] \hline  
$\mathrm{H}$  & $4.1\ (4.4)$
                  & $10.6\ (11.6)$
                  & $4.6\ (6.3)$   \\[2pt] \hline                   
$\mathrm{Total}$  & $17.6\ (21.4)$
                  & $42.1\ (51.9)$
                  & $19.3\ (37.5)$   \\[2pt] \hline
$\mathrm{MiniBN}$ & $19.5$
                  & $47.3$
                  & $19.4$   \\[2pt] \hline           
$\mathrm{Excess} $ & $42.6 \pm25.3$
                  & $82.2 \pm23.3$
                  & $21.5 \pm34.9$   \\[2pt] \hline                                                       
     \end{tabular}         
    \caption{$E_{QE}$ distribution of the \ncphoton\, events in the MiniBooNE 
neutrino run, comparing our estimate to the MiniBooNE estimate \cite{MiniBNoscsum}.}
    \label{tab:summary_EQE}
  \end{center}
\end{table}

\section*{Acknowledgments}
I acknowledge current support from the US Department of Energy under grant DE-FG02-93ER-40756, and is grateful to Tingjun Yang, Juan Nieves, and  Jan T. Sobczyk for helpful discussions during the workshop.

\end{document}